\documentclass[conference,a4paper]{IEEEtran}

\usepackage[utf8]{inputenc}
\usepackage[english]{babel}
\usepackage{cite}
\usepackage{hyperref}
\usepackage{graphicx}

\usepackage{amsmath}
\usepackage{bbm}

\usepackage{bm}
\renewcommand{\vec}[1]{\bm{#1}}
\newcommand{\indicator}[1]{\mathbbm{1}_{\{#1\}}}

\IEEEoverridecommandlockouts

\usepackage{tikz}

\newcommand\copyrighttext{%
  \footnotesize \textcopyright 2019 IEEE. Personal use of this material is permitted.
  Permission from IEEE must be obtained for all other uses, in any current or future
  media, including reprinting/republishing this material for advertising or promotional
  purposes, creating new collective works, for resale or redistribution to servers or
  lists, or reuse of any copyrighted component of this work in other works.
  DOI: \href{https://doi.org/10.1109/BlackSeaCom.2019.8812787}{10.1109/BlackSeaCom.2019.8812787}
  }
\newcommand\copyrightnotice{%
  \begin{tikzpicture}[remember picture,overlay]
  \node[anchor=south,yshift=10pt] at (current page.south) {\fbox{\parbox{\dimexpr\textwidth-\fboxsep-\fboxrule\relax}{\copyrighttext}}};
  \end{tikzpicture}%
}

\begin{document}

\title{Cloud-based Management of Energy-Efficient Dense IEEE 802.11ax Networks\thanks{The research has been carried out at IITP RAS and supported by the Russian Government (Contract No 14.W03.31.0019).}}

\author{\IEEEauthorblockN{
		Evgeny Khorov\IEEEauthorrefmark{1},
		Anton Kiryanov\IEEEauthorrefmark{1},
		Alexander Krotov\IEEEauthorrefmark{1}
	}
	\IEEEauthorblockA{ \IEEEauthorrefmark{1}Institute for Information Transmission Problems, Russian Academy of Sciences, Moscow, Russia\\
		Email: \{khorov, kiryanov, krotov\}@iitp.ru}
}

\maketitle

\copyrightnotice

\begin{abstract}
	During the last decade, the number of devices connected to the Internet by Wi-Fi has grown significantly. A high density of both the client devices and the hot spots posed new challenges related to providing the desired quality of service in the current and emerging scenarios. To cope with the negative effects caused by network densification, modern \mbox{Wi-Fi} is becoming more and more centralized. To improve network efficiency, today many new Wi-Fi deployments are under control of management systems that optimize network parameters in a centralized manner. In the paper, for such a cloud management system, we develop an algorithm which aims at maximizing energy efficiency and also keeps fairness among clients. For that, we design an objective function and solve an optimization problem using the branch and bound approach. To evaluate the efficiency of the developed solution, we implement it in the NS-3 simulator and compare with existing solutions and legacy behavior.
\end{abstract}

\section{Introduction}
In 2018, \mbox{Wi-Fi} traffic overtook the wired one. Traffic growth, as well as the increased number of devices and their density, raises new issues on how to increase the capacity of the network and provide high Quality of service for various traffic types. This task is significantly complicated by huge inter-network interference typical for nowadays deployments. In contrast to LTE systems, which share wireless spectrum in a centralized manner, \mbox{Wi-Fi} is based on random channel access and works distributively.

To improve performance of dense \mbox{Wi-Fi} networks, IEEE~802 LAN/MAN Standard committee is developing a new amendment to the \mbox{Wi-Fi} standard. This amendment, namely IEEE 802.11ax, introduces a palette of methods which can be jointly used to reduce interference, improve spectral efficiency and user experience in scenarios relevant to office and residential buildings, malls, airports, and stadiums \cite{khorov2019tutorial}.

\begin{figure}[tbp!]
	\centering
	\includegraphics[width=0.95\linewidth]{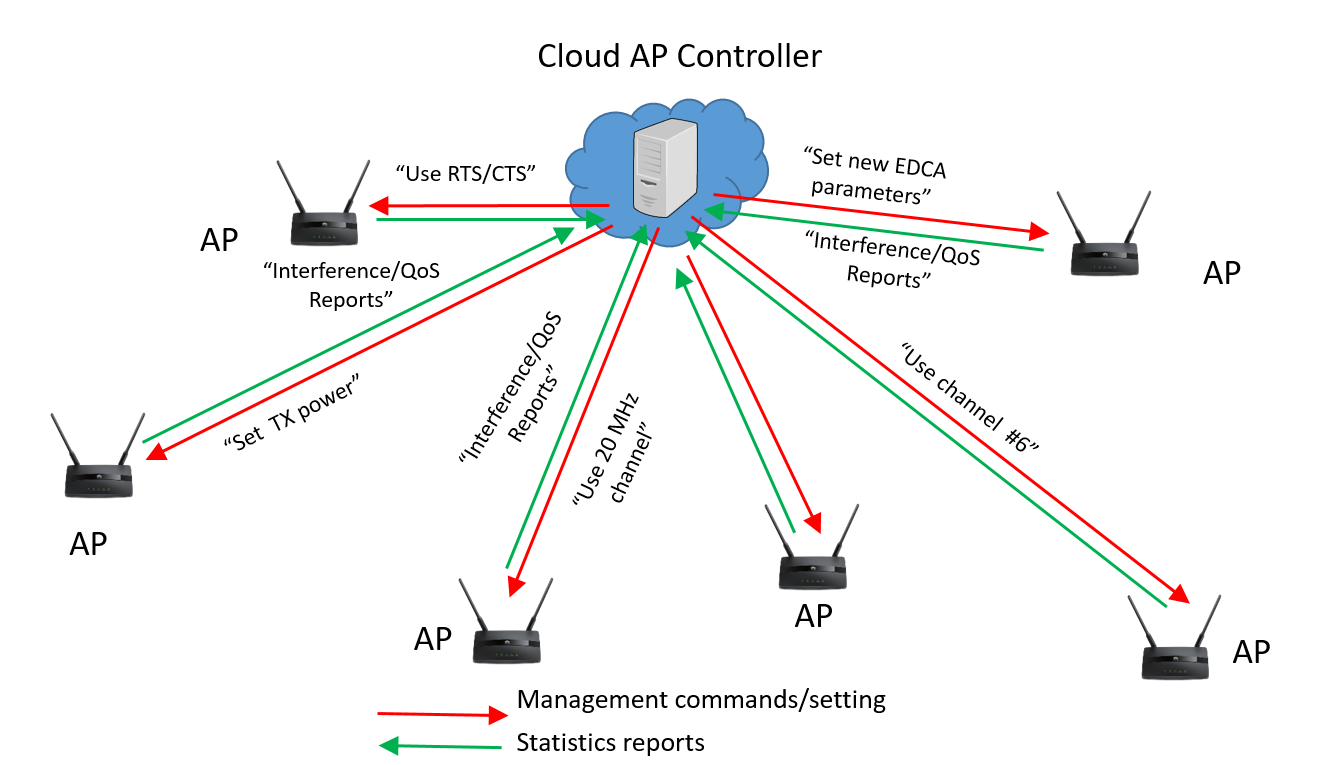}
	\caption{Cloud-based Wi-Fi Management System}
	\label{fig:arch}
\end{figure}

Apart from that, a recent \mbox{Wi-Fi} trend is large enterprise and home networks controlled by a single operator. In such deployments, it makes sense to introduce a centralized entity which can provide close coordination for neighboring access points (APs). Many vendors (including  HP/Aruba Networks, Cisco/Meraki, Quantenna Communications, and others) have advanced cloud infrastructure that can be used to control swarms of APs \cite{cloud}. Obviously, such cloud-based systems (see Fig.~\ref{fig:arch}) can provide real value for eliminating inter-cell interference, since they can have the solid picture of the interference in the network, and, thus, in theory, they can optimally schedule channel resources between various APs. A cloud-based management entity can control many \mbox{Wi-Fi} parameters like frequency channel, transmit power, sensitivity thresholds, and even time-division multiplexing, including those enabled by IEEE 802.11ax.

Besides, large \mbox{Wi-Fi} deployments may contain hundreds or even thousands of APs. In such a scenario, it becomes important to take into account the amount of energy consumed by each AP, and try to design energy-efficient solutions.

In this paper, we consider the problem of cooperation of \mbox{Wi-Fi} APs in order to achieve high network capacity with low energy consumption and fair channel allocation. Specifically, we state a global optimization problem and propose a centralized cloud-based algorithm which determines optimal operation parameters of controlled APs.

The rest of the paper is organized as follows. In Section~\ref{section:related}, we give a brief overview of prior art. In Section~\ref{section:problem}, we state the global optimization problem, Section~\ref{section:solution} contains the description of the proposed solution. Further, we evaluate the efficiency of the solution in Section~\ref{section:numerical} and give final remarks in Section~\ref{section:conclusion}.

\section{Related Work}
\label{section:related}

In the modern world, a high density of wireless networks and huge interference between them make centralized coordination of the networks more and more popular. It allows optimizing network performance and thus increasing total efficiency. While today's wireless networks are mainly optimized to provide high throughput, the growing OPEX of network operators including the payments for energy consumption may shift the paradigm in the near future. Because of the very high number of base stations and access points energy consumption becomes an essential issue for wireless networks.
To improve energy efficiency, various approaches can be used, including energy harvesting, improving hardware, network planning, and resource allocation~\cite{buzzi2016survey}.

In~\cite{zorzi1997energy}, energy efficiency is defined as the amount of data delivered through a link divided by the consumed energy. The authors of this paper consider a terminal having limited energy and compare the energy efficiency of various Automatic Repeat reQuest (ARQ)  protocols.

When optimizing energy efficiency, it is essential to add circuit power consumption $p_c$ to transmit power. Without taking this component into account, the maximum energy efficiency corresponds to the lowest transmission rate~\cite{li2011energy}.

Mentioned above papers consider only a single wireless link. The definition of energy efficiency has to be extended for systems with multiple transmitters and receivers. In ~\cite{miao2009interference}, it is done in the following way:
\begin{equation}
U = \sum_{i=1}^n U_i = \sum_{i=1}^n \frac{r_i}{p_i + p_c},
\label{eq:usum}
\end{equation}
\noindent where $U$ is the overall utility function, $U_i$ is the link utility function of link $i$, $n$ is the number of links, $r_i$ is the rate of link $i$, $p_i$ is the average transmit power at link $i$.
The major disadvantage of such a way is that the utility function represents the sum of energy efficiencies of individual links, while the network operator is interested in the total network energy consumption and energy efficiency which is different.

In~\cite{venturino2015energy}, the authors consider some other utility functions. In addition to the sum of energy efficiencies, an example of which is described above, they consider the product of energy efficiencies and so-called Global Energy Efficiency~(GEE). Global energy efficiency is defined as the sum of rates divided by the total power consumption of all devices. Fast algorithms are proposed to solve Sum-EE and Prod-EE maximization problems. For GEE maximization problem, the optimal solution is only found when interference is negligible compared to the constant background noise.

In this paper, we develop a global optimization algorithm which can be used to benchmark other approaches by comparing their results to the global maximum of utility functions.

The GEE maximization problem can be solved with existing mathematical methods based on the so-called polyblock algorithm~\cite{zappone2015energy}. However, this approach is known to converge very slowly when one or more variables are close to zero. While modeling real deployments, we often observed such cases. That is why we use another approach based on the branch-and-bound method that avoids this slow convergence~\cite{tuy2016convex}. Although being applied to solve the GEE problem\cite{zappone2017globally} in LTE networks, its applicability for \mbox{Wi-Fi} networks is not straightforward.

\mbox{Wi-Fi} networks impose additional restrictions on solutions of the described problem. Specifically, since \mbox{Wi-Fi} implements CSMA/CA, regulatory bodies put limits on the sensitivity threshold. An example of a solution to the GEE problem is shown in paper~\cite{blackseacom2018}, where an algorithm based on the branch-and-bound technique was proposed to allocate power in Wi-Fi networks dynamically. Even with a constant traffic load such an algorithm dynamically varies the transmit power and, thus, obtains higher efficiency. In this paper, we generalize the GEE metric to take both power consumption and fairness into account and to develop a global optimization algorithm for green Wi-Fi networks.

\section{Problem Statement}
\label{section:problem}
Let us consider $N$ established data links in a wireless network. We denote the link, the transmitter and the receiver of a link with the same index $i$. Suppose that data over the link $i$ is transmitted mostly in one direction. The total consumed power by link $i$ consists of a radiated power $p_i$ and a circuit power $p_{i,c}$. For simplicity, let the circuit power be the same for all devices, so $\forall i~ p_{i,c} = p_c = \text{const}$.
Let $0 \leq a_{ij} \leq 1$ be a pathloss between transmitter \(j\) and receiver \(i\),
and $0 \leq b_{ij} \leq 1$ be a pathloss between transmitter \(j\) and transmitter \(i\).
We suppose that all the receivers sense the signal of the corresponding transmitters, i.e.,
$\forall i~ a_{ii} > 0$. Apart from that, we assume $ \forall i~ b_{ii} = 0$.

The objective function~\eqref{eq:usum} has the following drawbacks. First, maximization of this function may lead to unfair resource sharing, which is obvious, e.g., if $p_c \gg p_i$. In this case, the objective function aims at maximizing total throughput which is known to be unfair. Second, it is more reasonable to optimize the energy efficiency of the whole network instead of the sum of energy efficiencies of individual links. Taking into account the notes above, in this paper, we propose to define the objective function as the ratio of the mean throughput to consumed power. In general case, the mean throughput is calculated as $U^{-1}\left(\frac{1}{N} \sum_{i=1}^N U(r_i)\right)$, where
\begin{equation}
\label{eq:utilit}
U(r_i) = \begin{cases}
\log(r_i), & \alpha = 1, \\
\frac{r_i^{1-\alpha}}{1 - \alpha}, & \alpha \ge 0, \alpha \ne 1.
\end{cases}
\end{equation}

\noindent For example, if $\alpha = 0$, we obtain the arithmetic mean. If $\alpha = 1$, we get the geometric mean.

So, the objective function $\hat U$ can be defined as follows:
\begin{equation}
\hat U = \frac{U^{-1}\left(\frac{1}{N} \sum_{i=1}^N U(r_i)\right)}{\sum_i^N (p_i + p_c)}.
\end{equation}

Note that if $\alpha = 0$, the problem is equivalent to the GEE maximization problem.

\section{Proposed Algorithm}
\label{section:solution}
\subsection{Utility Function Analysis}
Taking the logarithm of resulting objective function and representing it as the difference between two non-decreasing functions of data rates $\vec r$ we get
\begin{equation}
\label{eq:globutility}
\log \hat U (\vec r) = V (\vec r) - W (\vec r),
\end{equation}
where
\begin{equation}
V (\vec r)= \log U^{-1}\left(\frac{1}{N} \sum_i^N U(r_i)\right),
\end{equation}
and
\begin{equation}
W (\vec r) = \log \left(\sum_i^N (p_i + p_c)\right).
\label{eq:w}
\end{equation}

In~\cite[Lemma 2]{stefanyuk1967collective}, it is shown that for two SINR vectors $\gamma'$ and $ \gamma$, such that $\vec \gamma' \succeq \vec \gamma$ (i.e., each component of $\gamma'$ is not less than the corresponding component of $\gamma$), it follows that the corresponding vectors of transmit powers $\vec p' \succeq \vec p$. Since if $\vec r' \succeq \vec r$ then corresponding SINR vectors  $\vec \gamma' \succeq \vec \gamma$, we conclude that  \eqref{eq:w} is nondecreasing function of data rates.

Having represented~\eqref{eq:globutility} as the difference of monotonic functions, we can apply existing global optimization methods \cite[11.1.2 DM Functions and DM Constraints]{tuy2016convex}. According to \cite[Theorem 11.1]{tuy2016convex}, this problem can be represented as the optimization of a monotonic objective function subject to monotonic constraints. For this purpose, we introduce additional variable $w \in [-W(\vec b), -W(\vec a)]$, where $[W(\vec a), W(\vec b)]$ is target set of function $W$, and rewrite the problem statement as follows:
\begin{equation}
\max_r V(\vec r) + w
\label{eq:newobjective}
\end{equation}
subject to
\begin{equation}
W(\vec r) + w \leq 0.
\label{eq:newconstraint}
\end{equation}

The new objective function \eqref{eq:newobjective} is non-decreasing  function of $\vec r$ and $w$.

It should be mentioned that in addition to \eqref{eq:newconstraint} the following Wi-Fi receiver sensitivity constraints have to be considered:
\begin{equation}
\label{eq:cst}
\begin{aligned}
& \max_j b_{ij} p_j \le \hat c, \forall i \text{ such that } p_i>0; \\
& 0 \le p_i \le \hat p_i, \forall i.
\end{aligned}
\end{equation}

The first constraint reflects the carrier sense principle that shall be used by Wi-Fi devices since they operate in unlicensed bands. The transmission is allowed only when the received power is less than some threshold (i.e., the transmitter is not synchronized on a signal). The second constraint is the regulatory limitation on the total radiated power.

Note that rate $r_i$ is a non-decreasing function of SINR $\gamma_i$: $r_i = f(\gamma_i(\vec p))$, which can be estimated according to error rate models of available modulation and coding schemes, while SINR at receiver \(i\) is calculated as follows:
\begin{equation}
\label{eq:gamma}
\gamma_i(\vec p)= \frac{a_{ii} p_i}{n_i + \sum\limits_{j \neq i} a_{ij} p_j},
\end{equation}
where \(n_i\) is the thermal noise level at receiver $i$. Thus all the constraints can be represented for as
\begin{equation}
\label{eq:g}
g(\vec r, w) \leq 0,
\end{equation}
where $g(\vec r, w)$ is a non-decreasing function. It means that, if condition \eqref{eq:g} is not met at some point $(\vec r, w)$, then it is not satisfied at any other point $(\vec r', w')$ such that $(\vec r', w')  \succeq  (\vec r, w)$.

\subsection{Static Solution}
The solution of the described optimization problem can be found with the branch-and-bound algorithm. The main idea of the branch-and-bound algorithm is to iteratively split the search space $(\vec r, w)$ into box regions, calculate bounds on the objective function within these regions and reject regions that can not contain points with better objective function value than already found. The detailed explanation of the algorithm applied to a throughput optimization problem is described in~\cite{blackseacom2018}. The complexity of the algorithm and thus the processing time depends on the desired accuracy. In huge networks, we can significantly speed up calculations by limiting the number of iterations without noticeable losses in a performance gain.

\subsection{Dynamic Scheduling}
The described problem should be solved periodically because of at least two reasons. First, channel conditions, traffic load and activity of stations change over time. Second, in some cases, it is impossible to transmit data simultaneously in all overlapped Wi-Fi networks even with low transmit power. Thus, in the optimal solution, some transmissions will be forbidden. To avoid starvation, it is necessary to dynamically recalculate the solution taking into account the amount of traffic transmitted over different links.

Let $P$ and $R_i$ be the total energy spent from the beginning of the experiment and the total amount of data transmitted over the link $i$ from the beginning of the experiment. In this case, we can rewrite the objective function as follows:
\begin{equation}
\log \hat U = \log U^{-1}\left(\frac{1}{N} \sum_{i=1}^N U(R_i)\right) - \log P.
\label{ew:newobj}
\end{equation}

In case of dynamic selection, every time we want to maximize the utility function increment, i.e., to maximize the derivative. To achieve this, let us differentiate~\eqref{ew:newobj} taking into account~\eqref{eq:utilit}. For all $\alpha \geq 0$:
\begin{equation}
(\log \hat U)' = \frac{1}{\sum_{i=1}^N R_i^{1 - \alpha}} \sum_{i=1}^N \frac{R_i'}{R_i^\alpha} - \frac{1}{P}{P'}.
\end{equation}

To maximize the derivative, we use the approach that is described above.

\section{Numerical Results}
\label{section:numerical}

In this section, we provide and discuss the performance evaluation results of the proposed solution. We consider a square area 100 by 100 meters where we randomly place $10$ wireless clients (receivers). The clients are served by Wi-Fi APs (transmitters) the number of which varies from 1 to 30. The APs are arranged within the area in the way that maximizes the minimal distance between all pairs of APs and the walls.

We use the pathloss model from~\cite{simulation_scenarios}. So the signal attenuation at distance $d$ is calculated as follows:
\begin{equation*}
\label{eq:pathloss}
\begin{split}
d(r) &= 40.05 + 20 \log_{10}(f_c / 2.4) \\
&+ 20 \log_{10} (\min(r, 10)) + \indicator{r > 10} \cdot 35 \log_{10} 0.1 r,
\end{split}
\end{equation*}
where $f_c = 5.21$\,GHz, $\indicator{r > 10}$ is an indicator function equal to $1$ if $r > 10$, and $0$, otherwise.

\begin{figure}[!t]
	\centering
	\includegraphics[width=0.8\linewidth]{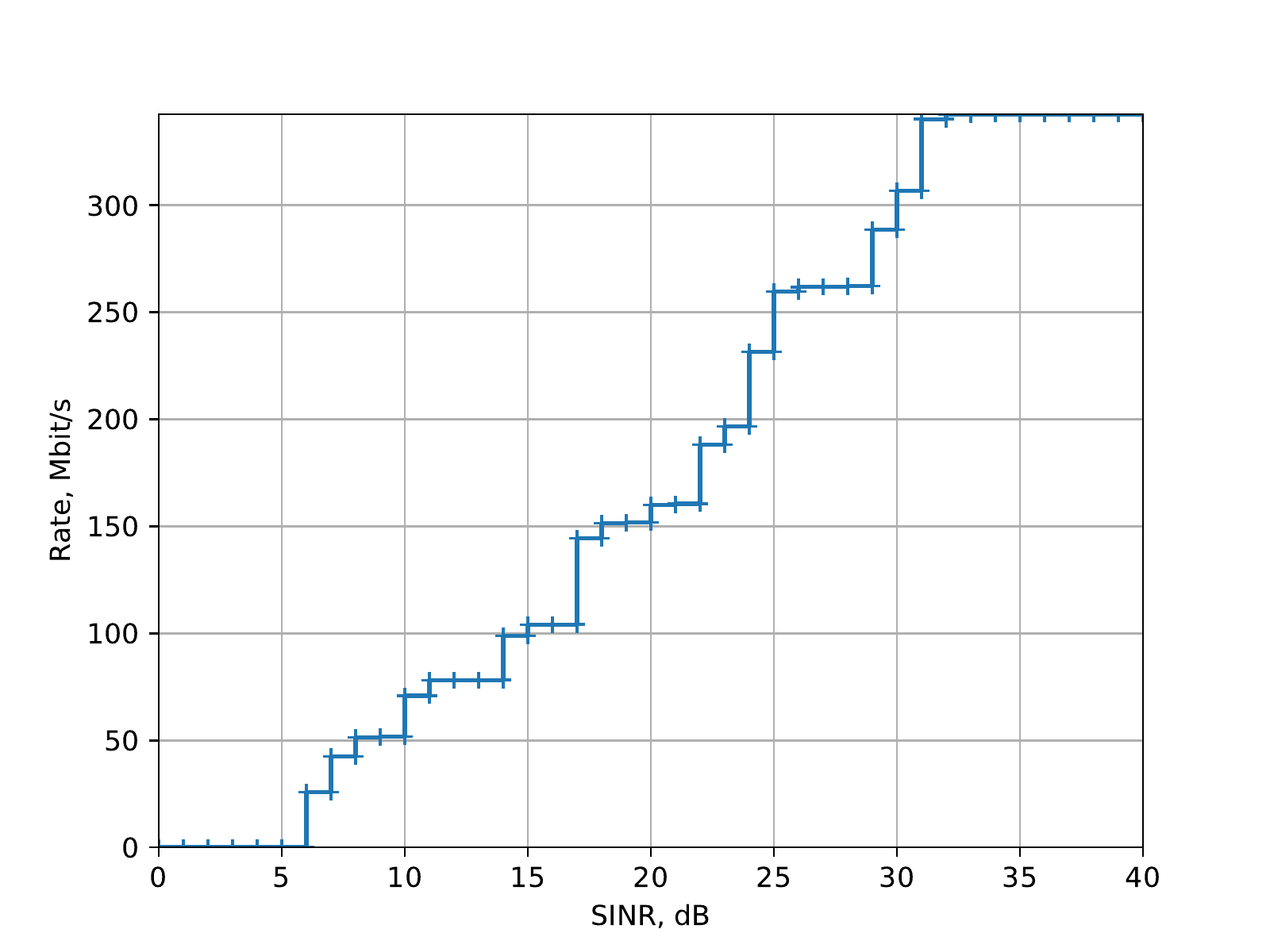}
	\caption{Data rate as a function of SINR}
	\label{fig:rate}
	\vspace{-1em}
\end{figure}

\begin{figure*}[h]
	\includegraphics[width=\linewidth]{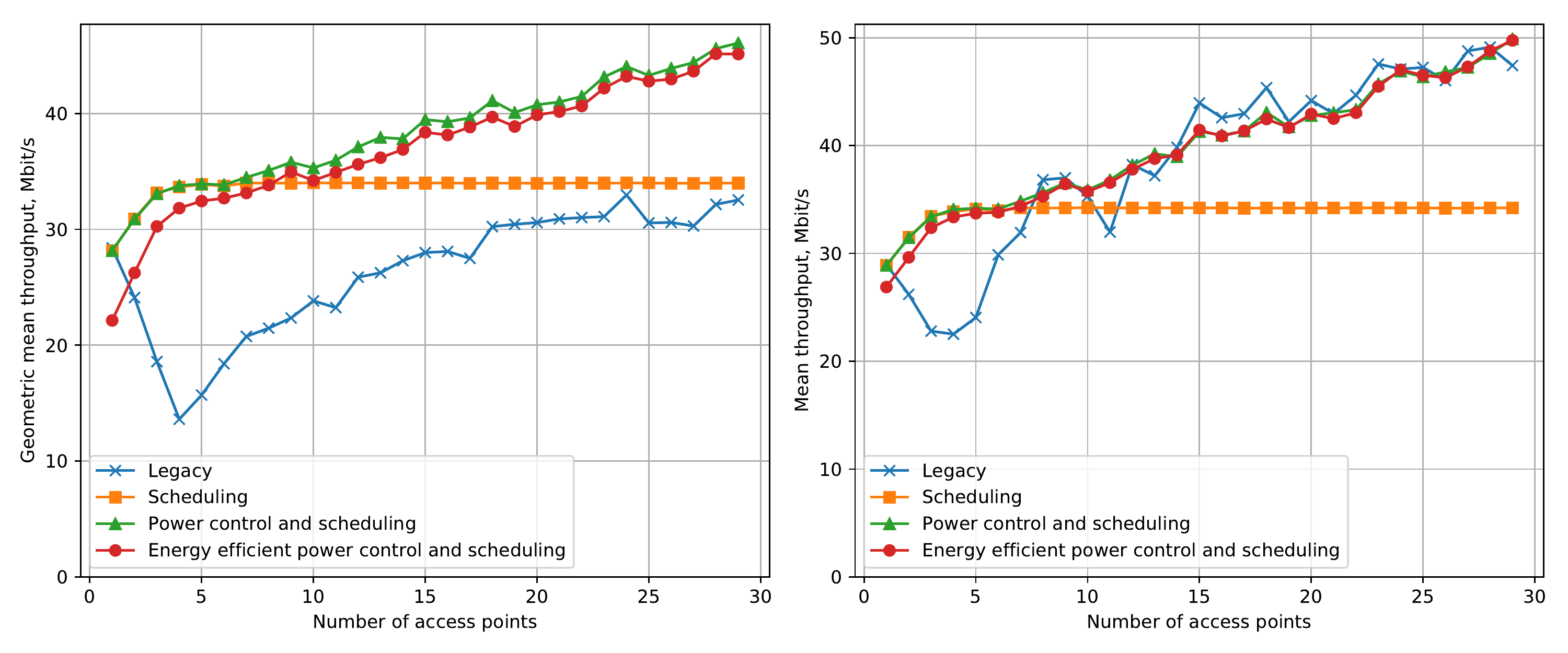}
	\caption{Throughput}
	\label{fig:thr}
	\vspace{-1em}
\end{figure*}

We solve the optimization problem described in Section~\ref{section:problem} for $\alpha = 1$. We have to specify how the data rate depends on SINR. Taking into account the simulation results obtained in NS-3~\cite{NS3} for the Minstrel rate selection algorithm, we approximate the dependence of rate on SINR by the following step function, see Fig.~\ref{fig:rate}.

\begin{table}[!t]
	\caption{Simulation parameters}
	\label{tab:params}
	\begin{center}
		\begin{tabular}{|l|c|}
			\hline
			Parameter & Value\\
			\hline
			AP location height, m & 3 \\
			Client location height, m & 1 \\
			Maximum transmit power, mW & 40 \\
			Noise, dBm/Hz & -174 \\
			Channel width, MHz & 80 \\
			Amplifier noise, dB & 7 \\
			Receiver Sensitivity, dBm & -96 \\
			Rate control algorithm & Minstrel~HT \\
			\hline
		\end{tabular}
	\end{center}
\end{table}

In the simulation, we compare the following solutions.
\begin{enumerate}
	\item Legacy. Legacy Wi-Fi behavior is modeled. No additional techniques or tuning is applied. All stations gain access to wireless medium according to Carrier Sense Multiple Access With Collision Avoidance (CSMA-CA).
	\item Scheduling. We schedule transmissions of APs in order to maximize geometric mean throughput. For that, we also solve the optimization problem described in~\cite{blackseacom2018}.
	\item Power control and scheduling. In addition to scheduling, we tune transmit power as described in \cite{blackseacom2018}.
	\item Energy-efficient power control and scheduling. The solution described in this work.
\end{enumerate}

\begin{figure}
	\includegraphics[width=\linewidth]{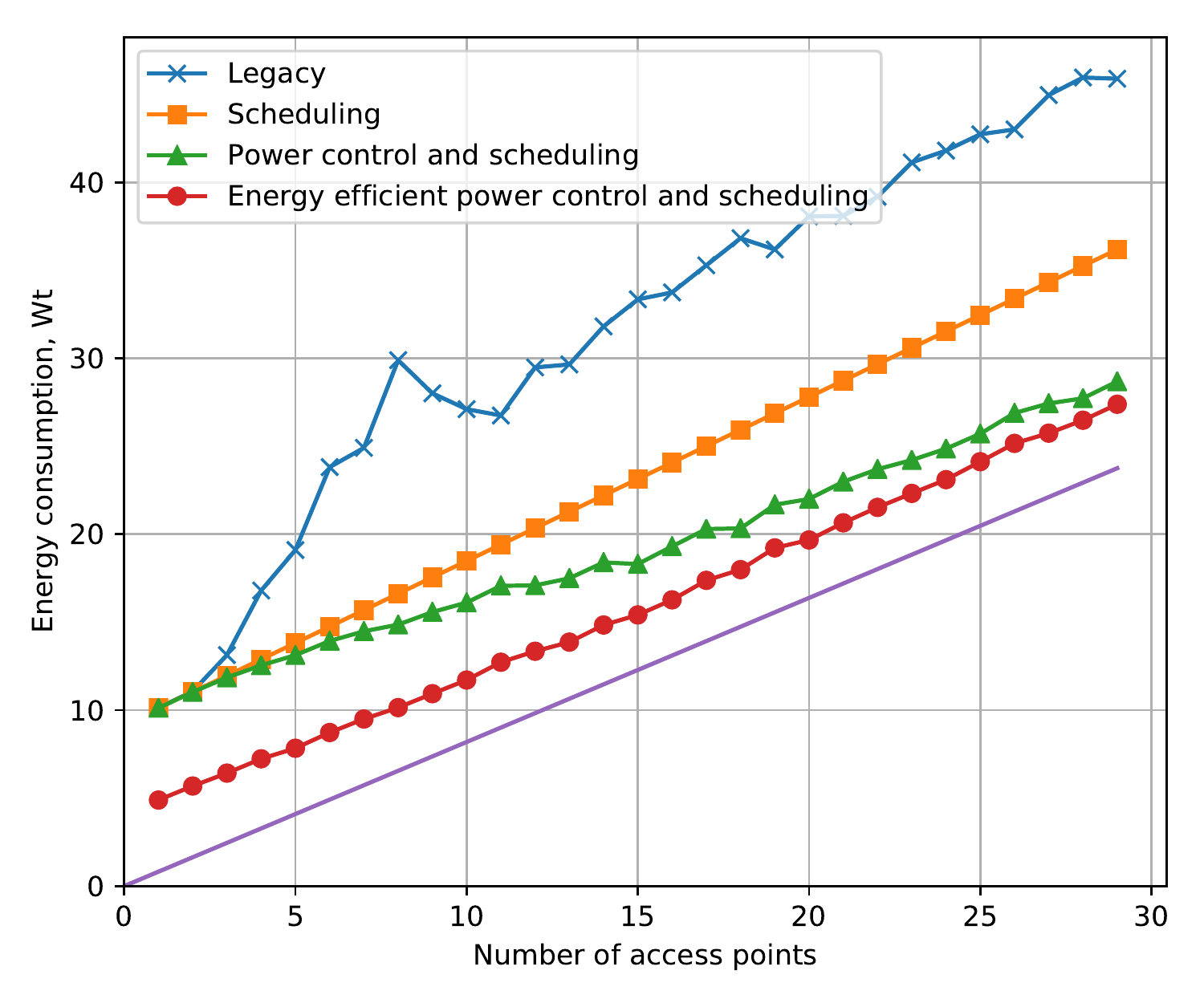}
	\caption{Energy consumption}
	\label{fig:energy}
	\vspace{-1em}
\end{figure}

\begin{figure}
	\includegraphics[width=\linewidth]{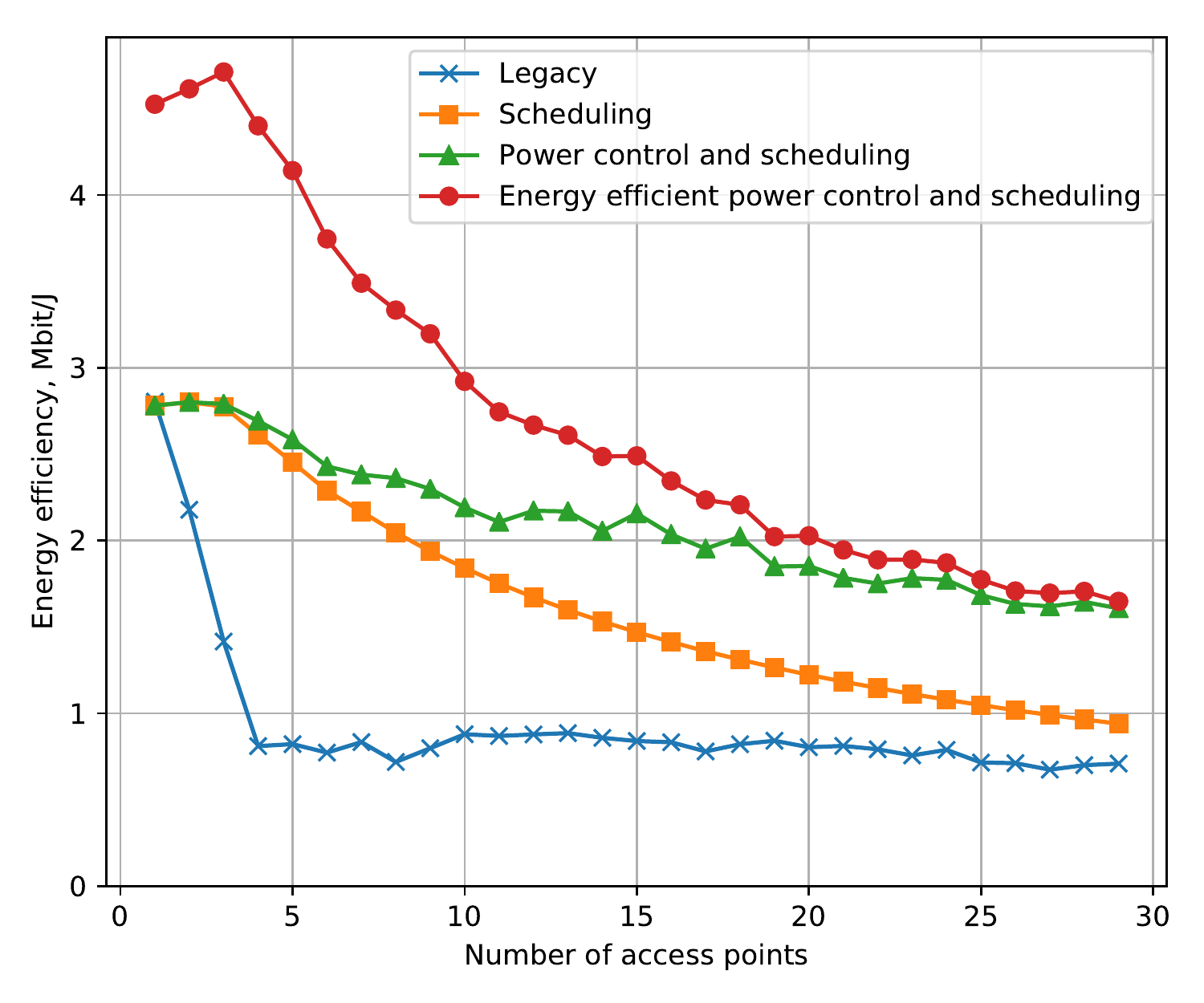}
	\caption{Energy efficiency}
	\label{fig:efficiency}
	\vspace{-1em}
\end{figure}

Fig.~\ref{fig:thr} shows the dependence of the mean throughput and the geometric mean throughput on the number of serving APs. For a single AP, the results of all solutions except for the one proposed in this paper coincide. This happens because for a single AP it does not matter how access to the medium is organized. In any case, this AP receives the whole channel. However, when trying to optimize energy efficiency, the solution proposed in this paper may sometimes reduce transmit power to save more energy which results in a bit lower throughput. For a legacy solution, increasing the number of serving APs does not bring any benefit. Instead, it leads to throughput degradation. This happens because of hidden stations and packet collisions which can be successfully eliminated by the AP coordination proposed in the paper. For more APs, throughput grows, because the average distance between the transmitter and the receiver decreases.  Although the legacy solution allows achieving a rather high mean throughput, it does not provide good enough fairness, which leads to smaller geometric mean throughput.

In case of scheduling, we observe the increase of throughput for a small number of APs, which is explained by the decrease of the average transmitter-receiver distance and thus the usage of faster Modulation and Coding Schemes (MCSs). After that, we see a constant throughput for any high number of APs, because in this case, at each time moment only one transmission occurs in the medium. Collisions inherent to the legacy solution are avoided, but no gain in throughput is achieved, because of the fixed transmit power. The curves for both the solutions with joint power control and scheduling (developed in \cite{blackseacom2018} and in this paper)   are quite close to each other (except for the case with a single AP). For both solutions, tuning transmit power increases the mean throughput from 30 Mbps up to 50 Mbps for a high number of APs. To understand the difference between these two solutions, let us look at Fig.~\ref{fig:energy} and \ref{fig:efficiency} with the total power consumption and energy efficiency. From these figures, we see that energy efficient power control allows substantially decreasing energy consumption while providing near the same throughput as our previous solution. In Fig.~\ref{fig:energy}, we also show a solid line which represents the minimal energy consumption in the network caused only by constant $p_c$ component. We observe that the results of the proposed algorithm are rather close to this lower limit (which may be reached only if the network transmits no data).

\section{Conclusion}
\label{section:conclusion}
In this paper, we propose a new algorithm to manage transmissions in dense Wi-Fi networks in order to enhance energy efficiency. The algorithm uses a branch-and-bound approach to maximize energy efficiency in a fair manner. To evaluate the performance of the designed algorithm, we implemented it in well-known NS-3 simulator. Obtained simulation results show up to 50\% increase in the energy efficiency comparing to our previously designed algorithm while providing similar mean and geometric mean throughput. As the next steps of the research, we are going to evaluate the efficiency of the designed solutions in a dynamically changing environment.

\bibliographystyle{IEEEtran}
\bibliography{biblio,main}

\end{document}